\documentclass[fleqn,10pt,a4paper]{article}
\usepackage{amsmath,amssymb,authblk,bm}
\usepackage{cite,epsfig,color,graphicx,subfigure}

\title{Plasma-based wakefield accelerators as sources of axion-like particles}

\author[1]{David A. Burton}
\author[2]{Adam Noble}
\affil[1]{Department of Physics\\ Lancaster University\\ Lancaster\\ LA1 4YB\\ UK}
\affil[2]{Department of Physics\\ SUPA and University of Strathclyde\\ Glasgow\\ G4 0NG\\ UK}

\begin{document}
\maketitle
\begin{abstract}
We estimate the average flux density of minimally-coupled axion-like particles generated by a laser-driven plasma wakefield propagating along a constant strong magnetic field. Our calculations suggest that a terrestrial source based on this approach could generate a pulse of axion-like particles whose flux density is comparable to that of solar axion-like particles at Earth. This mechanism is optimal for axion-like particles with mass in the range of interest of contemporary experiments designed to detect dark matter using microwave cavities.
\end{abstract}

\section{Introduction}
The next generation of high-intensity laser facilities is expected to provide a new avenue for probing quantum electrodynamics in parameter regimes that are currently inaccessible to high-energy particle colliders. Facilities such as ELI~\cite{eli} will offer interacting laser fields of such high intensity ($\sim 10^{23}\,{\rm W}\,{\rm cm^{-2}})$ that it will be possible to investigate the effective self-coupling of the electromagnetic field via virtual electron-positron pairs~\cite{dipiazza:2012}, and to explore the recoil of matter due to its own electromagnetic emission~\cite{burton:2014}. Furthermore, it is possible that a new perspective will be gained on long-standing problems in fundamental physics that include understanding the nature of dark matter. In particular, such facilities are expected to be capable of complementing established approaches for exploring the existence of light weakly-interacting axion-like particles (ALPs)~\cite{mendonca:2007, dobrich:2010, villalba:2013, villalba:2014}. Axions were first introduced in the 1970s as an elegant solution to the strong CP problem in QCD~\cite{peccei:1977, weinberg:1978, wilczek:1978}, but it was later realised that ALPs, which are light pseudo-scalar particles whose couplings to ordinary matter resemble those of the axion, naturally occur in string-inspired generalisations of the Standard Model~\cite{cicoli:2012}. The significance of the axion as a dark-matter candidate was first revealed during the early 1980s~\cite{preskill:1983, abbott:1983, dine:1983}, and this inspired the development of a wealth of experiments to search for axions and their ALP brethren. See Refs.~\cite{baker:2013, rosenberg:2015} for a recent summary of the most established programmes devoted to searching for axions and ALPs. It is also worth noting that the panoply of proposed experiments is continuing to grow; for example, it has been suggested that photonic band-gap structures~\cite{seviour:2014} and arrays of dielectric discs~\cite{caldwell:2017} could be fruitful for axion/ALP searches.

Investigations of the implications of quantum electrodynamics and non-Standard Model physics in the context of high-intensity laser experiments commonly use laser pulses as experimental probes. In tandem with such studies, it is also interesting to explore the behaviour of matter accelerated by electromagnetic fields that carry the imprint of the quantum vacuum and may also include effects due to new physics. Our particular interest here is on the paradigm of plasma-based wakefield acceleration~\cite{tajima:1979}, which has gained a prominent status in recent years as an effective method of efficiently accelerating charged particles. Landmark demonstrations of this mechanism~\cite{mangles:2004, geddes:2004, faure:2004} during the early years of the 21st century inspired the substantial worldwide effort currently devoted to harnessing plasma-based wakefield accelerators for practical purposes. The concept is particularly attractive because the electric fields in a plasma wave can be several orders of magnitude greater than those sustainable in standard radio-frequency accelerator cavities. The most prevalent schemes realised thus far employ the strong fields in the wake behind an intense laser pulse propagating through the plasma~\cite{schlenvoigt:2008}, although electron-driven wakefields have also been exploited for electron acceleration~\cite{blumenfeld:2007}. Furthermore, recent developments have focussed on proton-driven wakefields as a paradigm for efficiently accelerating leptons to TeV energies~\cite{assman:2014}.

The purpose of this article is to argue that the combination of a plasma wakefield accelerator and a constant strong magnetic field could offer an interesting source of ALPs for fundamental physics experiments. In general, a detailed investigation of the evolution of a laser-driven, or particle-driven, plasma wakefield requires intensive numerical simulation~{\color{black} \cite{esarey:2009}}. The most sophisticated numerical analyses of plasma wakefields are typically undertaken using $3$-dimensional particle-in-cell codes running on high-performance computer facilities, but reduced models and scaling arguments~{\color{black} \cite{pukhov:2002, lu:2007}} are also widely exploited for practical reasons associated with computational efficiency. Even though the wakefields in plasma-based particle accelerators are $3$-dimensional in nature, $1$-dimensional models play an important role and are widely used to obtain estimates of key quantities such as the acceleration gradient~{\color{black} \cite{esarey:2009}}. Our aim here is to obtain analytical expressions and numerical estimates for informing further study, and $1$-dimensional models are indispensible in this context.

For simplicity, throughout the following we will use Heaviside-Lorentz units with $c=1$, $\hbar=1$ unless otherwise stated.
\section{Field equations}
Minimally-coupled axion-like particles (ALPs) of mass $m_\Psi$ are described by a pseudo-scalar field $\Psi$ that satisfies the sourced Klein-Gordon equation:
\begin{equation}
\label{ALP_eqn}
\partial_t^2\Psi - \bm{\nabla}^2\Psi + m_\Psi^2 \Psi = -g {\bf E}\cdot{\bf B}
\end{equation}
where the constant $g$ is the ALP-photon coupling strength, ${\bf E}$ is the electric field and ${\bf B}$ is the magnetic field. The reaction of ${\bf E}$, ${\bf B}$ to the ALP field $\Psi$ is facilitated by the electromagnetic constitutive equations ${\bf D} = {\bf E} - g\Psi {\bf B}$, ${\bf H} = {\bf B} + g\Psi {\bf E}$ in the absence of material polarisation or effects due to the quantum vacuum; thus, the ALP field affects the electromagnetic field in a manner that is similar to a magnetoelectric medium. From the above perspective, conventional QCD axions are a class of minimally-coupled ALP whose parameters $g$, $m_\Psi$ lie within a band that straddles a particular line in the $\log(g) - \log(m_\Psi)$ plane. However, the results discussed here are applicable to any hypothetical pseudo-scalar particle that couples to the electromagnetic field in the above manner.

Effects due to quantum vacuum polarisation are readily introduced via an effective self-coupling of the electromagnetic field induced from a Lagrangian that can be expressed as a scalar-valued function ${\cal L}_{\rm EM}(X,Y)$ of the Lorentz invariants $X = |{\bf E}|^2 - |{\bf B}|^2$, $Y = 2{\bf E}\cdot{\bf B}$. In this case, the electromagnetic constitutive equations are
\begin{equation}
{\bf D} = 2\bigg(\frac{\partial{\cal L}_{\rm EM}}{\partial X}\, {\bf E} + \frac{\partial{\cal L}_{\rm EM}}{\partial Y}\, {\bf B}\bigg) - g\Psi {\bf B},\qquad{\bf H} = 2\bigg(\frac{\partial{\cal L}_{\rm EM}}{\partial X}\, {\bf B} - \frac{\partial {\cal L}_{\rm EM}}{\partial Y}{\bf E}\bigg) + g\Psi{\bf E}.
\label{DH_constitutive}
\end{equation}
The simplest choice, ${\cal L}_{\rm EM}=X/2$, corresponds to classical vacuum electromagnetism, whilst vacuum polarisation due to virtual electron-positron pairs can be accommodated by the weak-field $1$-loop Euler-Heisenberg Lagrangian~{\color{black}\cite{marklund:2006}}
\begin{equation}
\label{EH_leading_order}
\mathcal{L}_{\rm EM} = \frac{X}{2} + \frac{2\alpha^2}{45m_e^4}\left(X^2 +\frac{7}{4}Y^2\right)
\end{equation}
where $\alpha \approx 1/137$ is the fine-structure constant and $m_e$ is the electron mass. Another famous choice for ${\cal L}_{\rm EM}(X,Y)$ was introduced by Born and Infeld in an attempt to fix problems in quantum theory that arose due to the Coulomb singularity of the electron~\cite{born:1934}. Although interest in Born-Infeld theory waned soon after the development of renormalised QED, it was later rejuvinated because the Born-Infeld Lagrangian was shown to emerge from bosonic open string theory~\cite{fradkin:1985}. String-theoretic and other considerations have been used to motivate a wide range of non-linear theories of electromagnetism, and it is fruitful to examine the general implications of (\ref{DH_constitutive}).

A sufficiently short and intense laser pulse, or particle bunch, propagating through a plasma will leave a non-linear electron density wave in its wake. For computational simplicity, we will adopt the most widely used methodology for obtaining analytical estimates of the properties of the wake. Specifically, although one can include thermal effects in the analysis of the wake~\cite{burton:2010, thomas:2016}, we will represent the ordinary matter sector by a cold (i.e. pressureless) plasma. {\color{black} In any case, thermal effects are not expected to be significant for the parameter regime of interest here}. Furthermore, the motion of the ions due to the driving laser pulse or particle bunch is negligible in comparison to the motion of the electrons because the plasma ions are substantially more massive than the plasma electrons. Thus, we will neglect the motion of the ions.

The plasma electrons satisfy
\begin{equation}
\label{momentum_balance}
\partial_t {\bf p} + ({\bf u}\cdot \bm{\nabla}) {\bf p} = - e ({\bf E} + {\bf u}\times{\bf B})
\end{equation}
where $-e$ is the electron charge, ${\bf u}$ is the plasma electrons' $3$-velocity and
\begin{equation}
{\bf p} = \frac{m_e {\bf u}}{\sqrt{1-{\bf u}^2}}
\label{pu_relation}
\end{equation}
is their relativistic $3$-momentum. The plasma ions are are assumed to be uniformly distributed and static over the time and length scales of interest, and the electric charge density $\rho$ and electric current density ${\bf J}$ are
\begin{equation}
\label{sources}
\rho = \rho_0 + \rho_e,\qquad {\bf J} = \rho_e\,{\bf u}
\end{equation} 
where $\rho_e$ is the electron charge density and $\rho_0$ is the background ion charge density (a positive constant). The final ingredient that closes the system of field equations for $\Psi, {\bf E}, {\bf B}, {\bf u}, \rho_e$ are Maxwell's equations:
\begin{align}
\label{DH_eqns}
\bm{\nabla}\cdot{\bf D} = \rho,&\qquad \bm{\nabla}\times{\bf H} = {\bf J} + \partial_t{\bf D},\\
\label{EB_eqns}
\bm{\nabla}\cdot{\bf B} = 0,&\qquad \bm{\nabla}\times{\bf E} = -\partial_t{\bf B}.
\end{align}

Our interest lies in the properties of solutions to the above system that describe the behaviour of the ALP field $\Psi$ coupled to a non-linear electron density wave in a magnetised plasma. The wave propagates parallel to the applied magnetic field, which is assumed to be uniform with strength $B$. A well-established simple strategy for modelling wakefield accelerators is to focus on solutions in which all scalar fields and vector fields depend on $\zeta = z - v t$ only, where $\zeta$ is the phase of the wave, the constant $v$ is the phase speed of the wave and $0 < v < 1$. Furthermore, we will choose all vector fields to be proportional to $\hat{{\bf z}}$.

The above simplifications reduce the field equations to an amenable system of non-linear ODEs. To proceed, we note that (\ref{DH_eqns}), (\ref{EB_eqns}) reduce to
\begin{equation}
\label{D_ODEs}
D^\prime = \rho_0 + \rho_e,\qquad -vD^\prime + \rho_e u = 0,
\end{equation}
whilst (\ref{momentum_balance}), (\ref{ALP_eqn}) become
\begin{equation}
\label{momentum_ODE}
-vp^\prime +up^\prime = - eE
\end{equation}
and
\begin{equation}
\label{ALP_ODE}
v^2\Psi^{\prime\prime} - \Psi^{\prime\prime} + m_\Psi^2 \Psi = -gEB
\end{equation}
respectively, where ${\bf D} = D(\zeta)\,\hat{{\bf z}}$, ${\bf E} = E(\zeta)\,\hat{{\bf z}}$, ${\bf p} = p(\zeta)\,\hat{{\bf z}}$, ${\bf u} = u(\zeta)\,\hat{{\bf z}}$, $\Psi = \Psi(\zeta)$, $\rho_e = \rho_e(\zeta)$, whilst ${\bf B} = B\,\hat{{\bf z}}$ is constant. A prime indicates differentiation with respect to $\zeta$. The electric displacement $D$ is given in terms of $E$, $B$, $\Psi$ as follows:
\begin{equation}
D = 2\bigg(\frac{\partial{\cal L}_{\rm EM}}{\partial X}\, E + \frac{\partial{\cal L}_{\rm EM}}{\partial Y}\, B\bigg) - g\Psi B
\end{equation}
where (\ref{DH_constitutive}) has been used.

The next step is to encode the plasma electron momentum in a manner that is useful for describing a non-linear density wave. We express $p(\zeta)$ in terms of a dimensionless function $\xi=\xi(\zeta)$ as follows:
\begin{equation}
\label{p_encoding}
p = m_e \gamma (v\xi - \sqrt{\xi^2-1})
\end{equation}
where $\gamma=1/\sqrt{1-v^2}$ and $m_e \xi$ is the plasma electrons' relativistic energy in the inertial frame moving with velocity $v\,\hat{{\bf z}}$ with respect to the plasma ions (henceforth called the {\it wave frame}). Using (\ref{pu_relation}), (\ref{p_encoding}) it follows that
\begin{equation}
\label{u_encoding}
u = \frac{v\xi - \sqrt{\xi^2-1}}{\xi - v\sqrt{\xi^2 - 1}}
\end{equation}
and, since $\xi$ cannot be less than unity and $u$ is a monotonically decreasing function of $\xi$ for $\xi>1$ and $0 < v < 1$, it follows that the velocity of the electrons is always less than the phase speed of the wave (i.e. $u<v$) except in the critical case where $\xi=1$.

Motion during which $\xi$ approaches unity is highly non-linear; using (\ref{D_ODEs}) it follows that
\begin{equation}
\label{electron_density}
\rho_e = \frac{-v\rho_0}{v-u}
\end{equation}
and so the charge density $\rho_e$ diverges as $u\rightarrow v$. The profile of the electric field $E$ steepens accordingly. Using (\ref{momentum_ODE}), (\ref{p_encoding}), (\ref{u_encoding}) it follows that
\begin{equation}
\label{E_xi}
E = -\frac{m_e}{\gamma e} \xi^\prime
\end{equation}
and the quantity $m_e \xi / e$ emerges as an electrostatic potential in the wave frame.

The highly non-linear oscillation during which the plasma electron velocity $u$ grazes the phase speed $v$ of the wave exhibits the largest electric field that can be obtained for $0<v<1$ without losing the structure of the wave. Larger amplitude waves of the above regular form cannot be sustained; instead, experiments and detailed numerical simulation show that substantial numbers of electrons are drawn from the plasma, become trapped in the collapsing wave structure and are strongly accelerated. This effect is well established for classical vacuum electromagnetism (i.e. ${\cal L}_{\rm EM} = X/2$) and, from a perturbative perspective, it is expected to occur in the presence of new physics.

The above choices reduce (\ref{D_ODEs})-(\ref{E_xi}) to a pair of coupled non-linear ODEs for $\xi$, $\Psi$. The ODE for $\Psi$ is just the remnant (\ref{ALP_ODE}) of the ALP field equation, which can be written as
\begin{equation}
\label{ALP_ODE_simp}
\frac{\Psi^{\prime\prime}}{\gamma^2} - m_\Psi^2 \Psi = gEB
\end{equation}
where (\ref{E_xi}) is understood. However, it is fruitful to eschew the ODE for $\xi$ in favour of a first integral of the system. Substantial insight into the behaviour of the system is readily gained by combining (\ref{D_ODEs})-(\ref{E_xi}) to give
\begin{equation}
\label{first_integral_ODE}
\bigg[2\bigg(\frac{\partial{\cal L}_{\rm EM}}{\partial X} E^2 + \frac{\partial{\cal L}_{\rm EM}}{\partial Y} E B\bigg) - {\cal L}_{\rm EM} - \frac{1}{2}\bigg(\frac{\Psi^{\prime 2}}{\gamma^2} - m_{\Psi}^2\Psi^2\bigg) - \frac{m_e\gamma \rho_0}{e}(v\sqrt{\xi^2-1}-\xi)\bigg]^\prime = 0,\,\,\,\,
\end{equation}
which can also be obtained using a manifestly covariant approach by invoking total stress-energy-momentum balance~\cite{burton:2016}. The pair (\ref{ALP_ODE_simp}), (\ref{first_integral_ODE}) of non-linear ODEs for $\xi$, $\Psi$ (with (\ref{E_xi}) understood), exhibit oscillatory solutions if an appropriate choice for ${\cal L}_{\rm EM}(X,Y)$ is made. Physically admissible choices for ${\cal L}_{\rm EM}(X,Y)$ lead to oscillatory solutions that can be smoothly transformed to those of ${\cal L}_{\rm EM} = X/2$ by taking an appropriate limit. An example of an admissible theory is given by (\ref{EH_leading_order}), which is perturbatively connected to classical vacuum electromagnetism by the fine-structure constant $\alpha$. Born-Infeld electrodynamics also falls into the category of admissible theories~\cite{burton:2011}.
\section{A novel source of axion-like particles}
\label{sec:ALP_flux}
An initial assessment of the effectiveness of plasma-based wakefields as sources of ALPs in the laboratory follows directly from (\ref{ALP_ODE_simp}), (\ref{first_integral_ODE}). The central result of this article is an analytical expression for the cycle-averaged ALP flux density $N_\Psi$ associated with the critical oscillatory solution to (\ref{ALP_ODE_simp}), (\ref{first_integral_ODE}). Use of the critical solution ensures that the source in (\ref{ALP_ODE_simp}) is optimal because, from a perturbative perspective, it has the largest amplitude electric field for given values of the parameters $g$, $m_\Psi$, $B$, $\rho_0$, $v$. The superposition of the critical wakefield and the applied magnetic field generates the largest amplitude ALP field at lowest order in the ALP-photon coupling strength $g$.

The ALP field $\Psi$ is static in the wave frame, and it follows that the relativistic energy of a single ALP is $\gamma m_\Psi$. Thus, $N_\Psi$ is
\begin{equation}
\label{av_N_Psi_final_c=1}
N_\Psi = \frac{\langle P_\Psi \rangle}{\gamma m_\Psi}
\end{equation}
where $P_\Psi=v \Psi^{\prime 2}$ is the $z$-component of the ALP energy flux density ${\bf P}_\Psi$,
\begin{equation}
\label{ALP_energy_flux_density}
{\bf P}_\Psi = -\partial_t \Psi\,\bm{\nabla}\Psi = v \Psi^{\prime 2}\,{\hat{\bf z}},
\end{equation} 
and $\langle P_\Psi \rangle$ is $P_\Psi$ averaged over an oscillation;
\begin{equation}
\label{ALP_energy_flux_average}
\langle P_\Psi \rangle = \frac{1}{T} \int^{T}_{0} P_\Psi\,dt
\end{equation}
with $T$ the period of the oscillation.

A minimally-coupled ALP field generated by the superposition of a periodic electric field and a constant magnetic field has the same period as the electric field. We express $\xi$ as
\begin{equation}
\label{xi_fourier}
\xi(\zeta) = \sum\limits^\infty_{n=-\infty} \xi_n \exp\bigg(2\pi i n \frac{\zeta}{l}\bigg) 
\end{equation}
where $l=vT$ is the wavelength of the electric field and, since $d\zeta = - v dt$ when $z$ is constant, (\ref{ALP_energy_flux_average}) can be written as
\begin{equation}
\label{average_P_Psi}
\langle P_\Psi \rangle = \frac{v}{l} \int^l_0 \Psi^{\prime 2}\,d\zeta. 
\end{equation}
The ALP field $\Psi$ has the form
\begin{equation}
\label{Psi_fourier}
\Psi(\zeta) = \sum\limits^\infty_{n=-\infty} \Psi_n \exp\bigg(2\pi i n \frac{\zeta}{l}\bigg) 
\end{equation}
where
\begin{equation}
\label{Psi_n}
\Psi_n = \frac{gBl\gamma}{4\pi^2 n^2 + m_\Psi^2 \gamma^2 l^2} \frac{m_e}{e} 2\pi i n \xi_n
\end{equation}
follows from (\ref{E_xi}), (\ref{ALP_ODE_simp}), (\ref{xi_fourier}), (\ref{Psi_fourier}).  Using (\ref{average_P_Psi}), (\ref{Psi_fourier}), (\ref{Psi_n}) it follows that
\begin{equation}
\label{P_Psi_xi_n}
\langle P_\Psi \rangle = v \sum\limits_{n=-\infty}^{\infty} \frac{16\pi^4 n^4 g^2 B^2 \gamma^2}{(4\pi^2 n^2 + m_\Psi^2 \gamma^2 l^2)^2} \frac{m_e^2}{e^2} |\xi_n|^2.
\end{equation}

A closed-form approximation to $\langle P_\Psi \rangle$ is readily obtained for the oscillation during which the amplitude of the electric field has its largest possible value. If the contribution to $\langle P_\Psi \rangle$ due to quantum vacuum polarisation and the back-reaction of the ALP field is neglected then
\begin{equation}
\label{xi_prime}
\frac{d\xi}{d\zeta}= \sqrt{2 \gamma^3} \omega_p (v\sqrt{\xi^2-1} - \xi + 1)^{1/2}
\end{equation}
follows from (\ref{first_integral_ODE}), where $\omega_p=\sqrt{e\rho_0/m_e}$ is the plasma frequency. Equation (\ref{xi_prime}) is obtained from (\ref{first_integral_ODE}) by making the substitutions ${\cal L}_{\rm EM} \mapsto X/2$, $\Psi \mapsto 0$ and demanding that $\xi=1$ when $d\xi/d\zeta=0$. The electric field $E$ has been eliminated using (\ref{E_xi}) during the passage from (\ref{first_integral_ODE}) to (\ref{xi_prime}).

Using (\ref{xi_prime}), the phase $\zeta$ of the wave can be written in terms of $\xi$ as
\begin{equation}
\label{zeta_integral}
\zeta(\xi) = \frac{1}{\sqrt{2\gamma^3}\omega_p} \int^\xi_1 \frac{1}{(v\sqrt{\chi^2 - 1} - \chi + 1)^{1/2}}\,d\chi
\end{equation}
where, for convenience, $\zeta(1)=0$ has been chosen. In practice, the Lorentz factor $\gamma$ of the wave satisfies $\gamma \gg 1$; for example, the Lorentz factor of a laser-driven plasma wakefield is in the range $10 - 100$ and the Lorentz factor of an electron-driven plasma wakefield is considerably larger ($\gamma \sim 10^5$). Thus, the approximation method that we will use to evaluate (\ref{zeta_integral}) is adapted to the large $\gamma$ regime. Introducing the scaled variables $\bar{\xi}=\xi/\gamma^2$, $\bar{\chi}=\chi/\gamma^2$ leads to
\begin{equation}
\label{xi_integral}
\zeta(\xi) = \frac{1}{\sqrt{2\gamma} \omega_p} \int^{\bar{\xi}}_{\gamma^{-2}} \bigg[\sqrt{\bigg(1-\frac{1}{\gamma^2}\bigg)\bigg(\bar{\chi}^2 - \frac{1}{\gamma^4}\bigg)} - \bar{\chi} + \frac{1}{\gamma^2}\bigg]^{-1/2}\,d\bar{\chi}
\end{equation} 
where $v=\sqrt{1-\gamma^{-2}}$ has been used to fully exhibit the $\gamma$ dependence of the integrand. However,
\begin{equation}
\sqrt{\bigg(1-\frac{1}{\gamma^2}\bigg)\bigg(\bar{\chi}^2-\frac{1}{\gamma^4}\bigg)} - \bar{\chi} + \frac{1}{\gamma^2} = \frac{1}{\gamma^2}\bigg(1-\frac{\bar{\chi}}{2}\bigg) + {\cal O}(\gamma^{-4})
\end{equation}
and it follows that the dominant behaviour of (\ref{zeta_integral}) when $\gamma \gg 1$ is
\begin{align}
\notag
\zeta &\approx \sqrt{\frac{\gamma}{2}} \frac{1}{\omega_p} \int^{\bar{\xi}}_0 \frac{1}{\sqrt{1-\bar{\chi}/2}}\,d\bar{\chi}\\
\label{zeta_approx}
& = \sqrt{\frac{\gamma}{2}}\frac{4}{\omega_p}\bigg(1-\sqrt{1-\frac{\bar{\xi}}{2}}\bigg).
\end{align}
Evaluating the period $l$ of the oscillation,
\begin{equation}
l = 2\zeta\big|^{\xi = \gamma^2 (1+v^2)}_{\xi = 1},
\end{equation}
using the approximate expression (\ref{zeta_approx}) for $\zeta$ gives
\begin{equation}
\label{l_approx}
l \approx 2\zeta\big|^{\bar{\xi} = 2}_{\bar{\xi}=0} = \frac{4\sqrt{2\gamma}}{\omega_p}
\end{equation}
when $\gamma \gg 1$.
Inverting (\ref{zeta_approx}) to find $\xi(\zeta)$ results in
\begin{equation}
\label{xi_approx}
\xi(\zeta) \approx 8\gamma^2 \frac{\zeta}{l}\bigg(1-\frac{\zeta}{l}\bigg) \quad \mbox{for $0 \le \zeta \le l$}
\end{equation}
where (\ref{l_approx}) has been used and the criterion $\gamma \gg 1$ is understood. An approximation to the coefficients $\xi_n$ in (\ref{xi_fourier}) follows immediately from (\ref{xi_approx}):
\begin{align}
\notag
\xi_n &= \frac{1}{l} \int^l_0 \exp(-2\pi i n \zeta/l)\,\xi(\zeta)\,d\zeta\\
\label{xi_n_approx}
&\approx
\begin{cases}
\displaystyle\frac{4\gamma^2}{3}\quad&\mbox{for $n=0$}\\
\displaystyle-\frac{4\gamma^2}{\pi^2 n^2}\quad&\mbox{for $n\neq 0$}
\end{cases}
\end{align}
and so (\ref{P_Psi_xi_n}) yields
\begin{equation}
\label{av_P_Psi_approx}
\langle P_\Psi \rangle \approx v g^2 B^2 \frac{m_e^2}{e^2} \frac{16\gamma^6}{\pi^4}\sum\limits_{n \neq 0} \frac{1}{(n^2 + s^2)^2}\quad\mbox{for $\gamma \gg 1$}
\end{equation}
where
\begin{align}
\label{s_exact}
s &= \frac{m_\Psi \gamma l}{2\pi}\\
\label{s_def}
&\approx \frac{m_\Psi}{\omega_p}\frac{2\sqrt{2}}{\pi}\gamma^{3/2}\quad\mbox{for $\gamma \gg 1$}
\end{align}
is obtained from (\ref{l_approx}). Since
\begin{equation}
\sum\limits_{n \neq 0} \frac{1}{(n^2 + s^2)^2} = \frac{\pi[\pi{\rm coth}^2(\pi s) s - \pi s + \coth(\pi s)]}{2 s^3} - \frac{1}{s^4}
\end{equation}
the result (\ref{av_P_Psi_approx}) can be expressed in closed form as follows:
\begin{equation}
\label{av_P_Psi_final}
\langle P_\Psi \rangle \approx \frac{\hbar c^4}{\mu_0^2} \frac{g^2 B^2 m_e^2}{e^2} \frac{16\gamma^6}{\pi^4}
\bigg\{\frac{\pi[\pi{\rm coth}^2(\pi s) s - \pi s + \coth(\pi s)]}{2 s^3} - \frac{1}{s^4}\bigg\}
\quad\mbox{for $\gamma \gg 1$}
\end{equation}
where
\begin{equation}
s = \frac{m_\Psi c^2}{\hbar \omega_p}\frac{2\sqrt{2}}{\pi}\gamma^{3/2}
\end{equation}
and $\hbar$, $c$, $\mu_0$ have been explicitly restored. The multiplicative factor $v$ in (\ref{av_P_Psi_approx}) has been replaced by $c$ in (\ref{av_P_Psi_final}) for consistency with the approximations used to obtain (\ref{av_P_Psi_approx}). Thus, the cycle-averaged flux density $N_\Psi$ of ALPs generated by the superposition of the wakefield and the applied magnetic field is
\begin{equation}
\label{av_N_Psi_final}
N_\Psi = \frac{\langle P_\Psi \rangle}{\gamma m_\Psi c^2}.
\end{equation}
\begin{figure}
\subfigure[]{
\begin{minipage}[t]{0.5\textwidth}
\centering\includegraphics[scale=0.6]{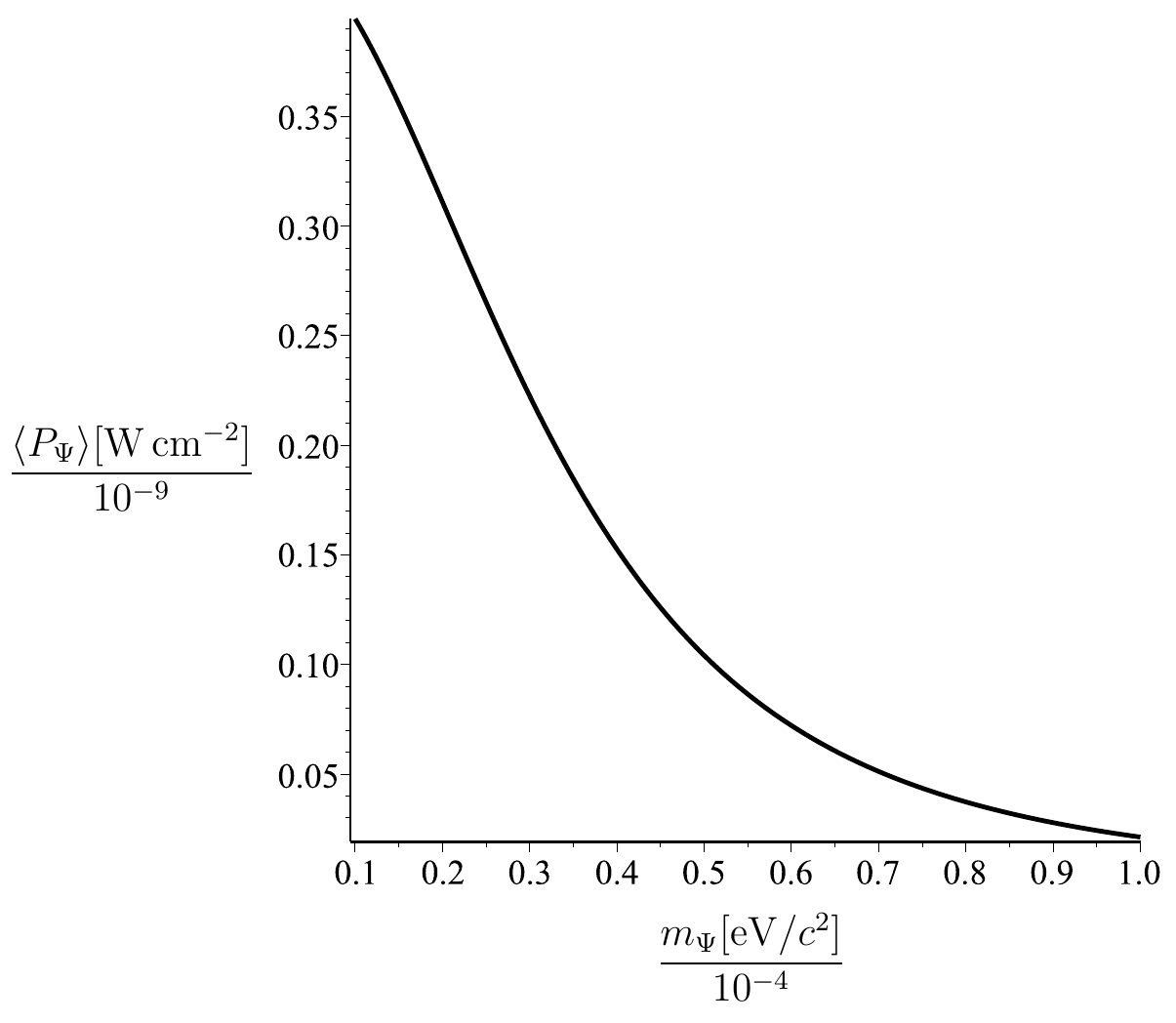}
\label{fig:P_vs_m}
\end{minipage}
}
\subfigure[]{
\begin{minipage}[t]{0.5\textwidth}
\centering\includegraphics[scale=0.6]{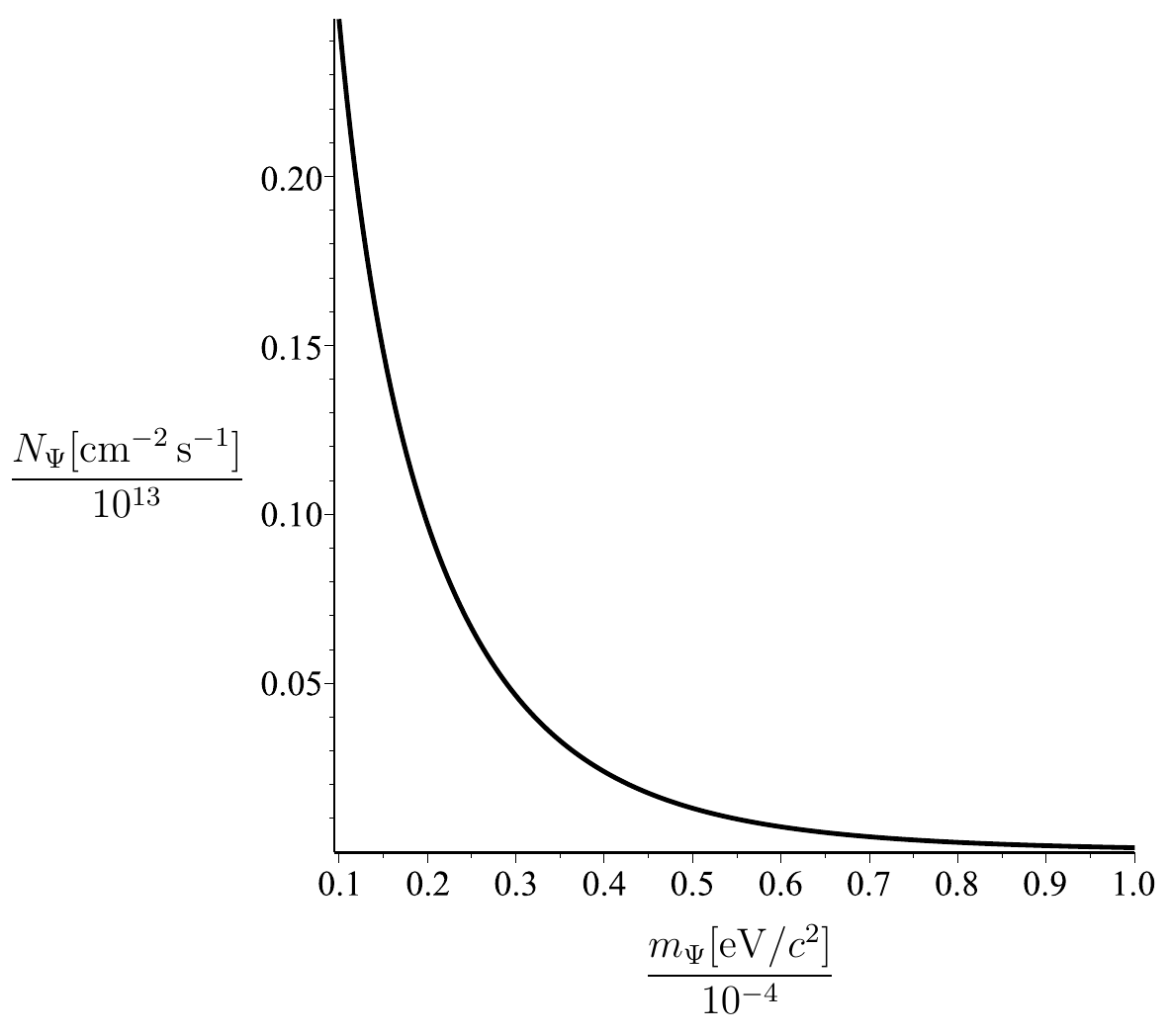}
\label{fig:N_vs_m}
\end{minipage}
}
\caption{The average ALP energy flux density (\ref{av_P_Psi_final}) and average ALP flux density (\ref{av_N_Psi_final}) versus the ALP mass $m_\Psi$ for the parameters $\omega_p = 2\pi\times 10^{13}\,{\rm rad}\,{\rm s}^{-1}$, $\gamma = 100$, {\color{black} $B = 35\,{\rm T}$, $g = 0.66 \times 10^{-10}\,{\rm GeV}^{-1}$}.}
\end{figure}

{\color{black} For practical purposes, the Lorentz factor $\gamma$ of the wake is commonly expressed in terms of the properties of the laser pulse driving the wake~\cite{esarey:2009}. Identifying the phase velocity $v$ of the wake with the group velocity of a laser pulse modelled using the dispersion relation $\omega_0^2 = c^2 k^2 + \omega_p^2$ from linear theory gives $\gamma=\omega_0/\omega_p$. Extensive investigation in the 1-dimensional and 3-dimensional non-linear regimes shows $\gamma \propto \omega_0/\omega_p$~\cite{lu:2007, schroeder:2011} and, in the former case, the coefficient of proportionality is sensitive to the structure and intensity of the laser pulse~\cite{schroeder:2011}.}

Representative parameters for a laser-driven plasma wakefield accelerator are $\omega_p \sim 2\pi\times 10^{13}\,{\rm rad}\,{\rm s}^{-1}$, $\gamma \sim 100$, whilst {\color{black} $B \sim 35\,{\rm T}$} is characteristic of the {\color{black} strongest solenoid magnets} available in the laboratory~\cite{NHMFL}. Typical estimates for $g$, $m_\Psi$ in the domain of interest can be obtained by appealing to the results of searches using the CERN Axion Solar Telescope (CAST)~{\color{black}\cite{CAST:2017}}. CAST has excluded ALPs with coupling strength {\color{black} $g \gtrsim 0.66 \times 10^{-10}\,{\rm GeV}^{-1}$} that satisfy $m_\Psi \lesssim 0.02\,{\rm eV}/c^2$.

A graph of $\langle P_\Psi \rangle$ versus $m_\Psi$ is shown in Fig.~\ref{fig:P_vs_m} where $\omega_p = 2\pi\times 10^{13}\,{\rm rad}\,{\rm s}^{-1}$, $\gamma = 100$, {\color{black} $B = 35\,{\rm T}$}, {\color{black} $g = 0.66 \times 10^{-10}\,{\rm GeV}^{-1}$}. Further investigation of (\ref{av_P_Psi_final}) shows that $\langle P_\Psi \rangle$ tends towards {\color{black} $4.3 \times 10^{-10}\,{\rm W}\,{\rm cm^{-2}}$} for $m_\Psi \lesssim 10^{-5}\,{\rm eV}/c^2$, which coincides with the narrow mass range $10^{-6}\,{\rm eV}/c^2\lesssim m_\Psi \lesssim 2\times 10^{-5}\,{\rm eV}/c^2$ in which the Axion Dark Matter Experiment (ADMX) operates~\cite{rosenberg:2015}. Fig.~\ref{fig:N_vs_m} shows the corresponding average ALP flux density $N_\Psi$ versus $m_\Psi$ given by (\ref{av_N_Psi_final}). For comparison, ALPs emitted by the Sun due to the Primakoff process are expected to have a flux density of $g_{10}^2\, 3.75\times 10^{11}\,{\rm cm}^{-2}\,{\rm s}^{-1}$ at Earth, where $g_{10} = g\,10^{10}\,{\rm GeV}$~\cite{andriamonje:2007}. If {\color{black} $m_\Psi \lesssim 1.8 \times 10^{-4} \,{\rm eV}/c^2$} then the flux density of ALPs generated by the plasma wakefield is greater than the solar ALP flux witnessed on Earth.
\begin{figure}
\subfigure[]{
\begin{minipage}[t]{0.5\textwidth}
\centering\includegraphics[scale=0.6]{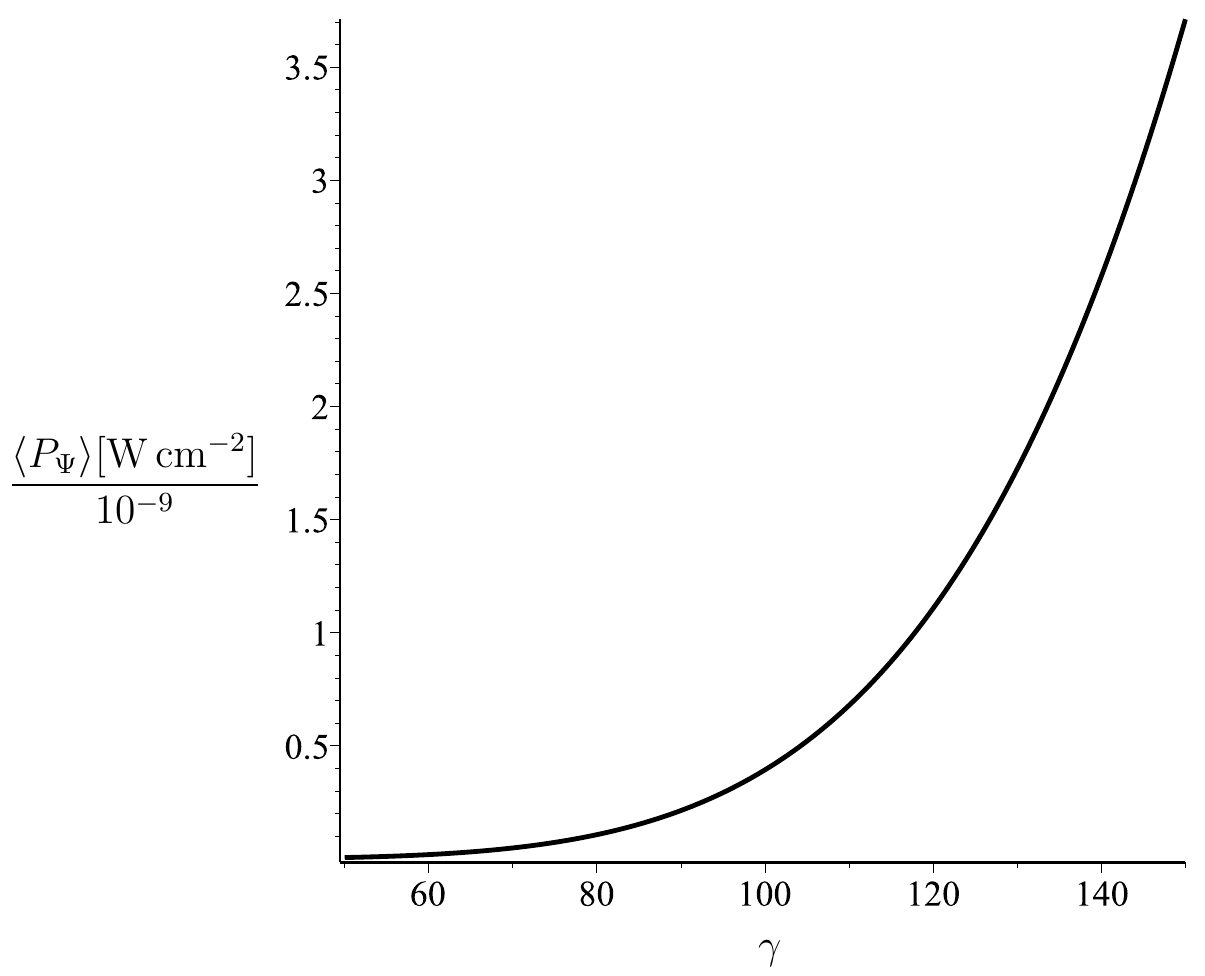}
\label{fig:P_vs_gamma_low_m}
\end{minipage}
}
\subfigure[]{
\begin{minipage}[t]{0.5\textwidth}
\centering\includegraphics[scale=0.6]{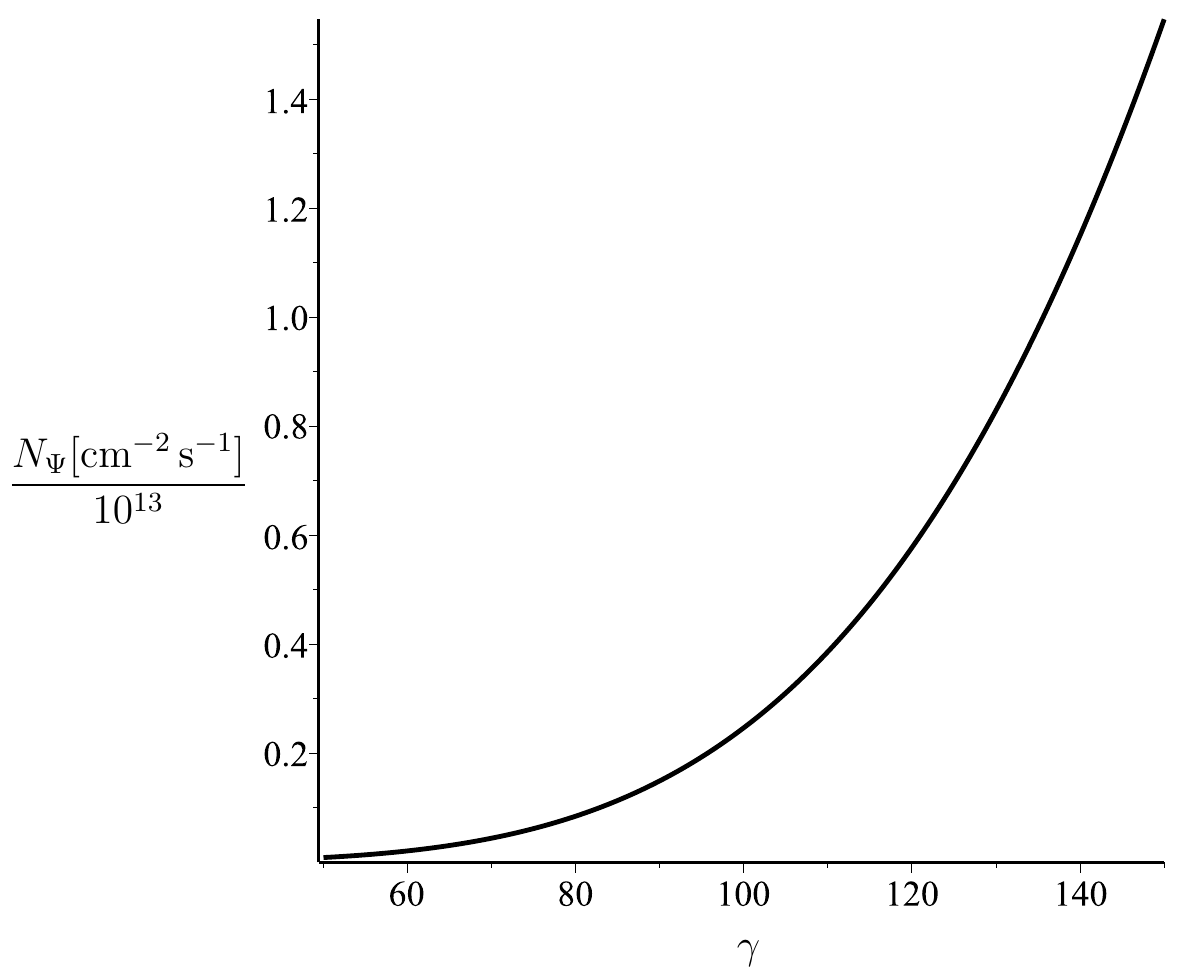}
\label{fig:N_vs_gamma_low_m}
\end{minipage}
}
{\color{black}\caption{\label{fig:fluxes_vs_gamma_low_m}The average ALP energy flux density (\ref{av_P_Psi_final}) and average ALP flux density (\ref{av_N_Psi_final}) versus the Lorentz factor $\gamma$ of the wake for the parameters $\omega_p = 2\pi\times 10^{13}\,{\rm rad}\,{\rm s}^{-1}$, $B = 35\,{\rm T}$, $g = 0.66 \times 10^{-10}\,{\rm GeV}^{-1}$, $m_\Psi = 10^{-5}\,{\rm eV}/c^2$.}}
\end{figure}
\begin{figure}
\subfigure[]{
\begin{minipage}[t]{0.5\textwidth}
\centering\includegraphics[scale=0.6]{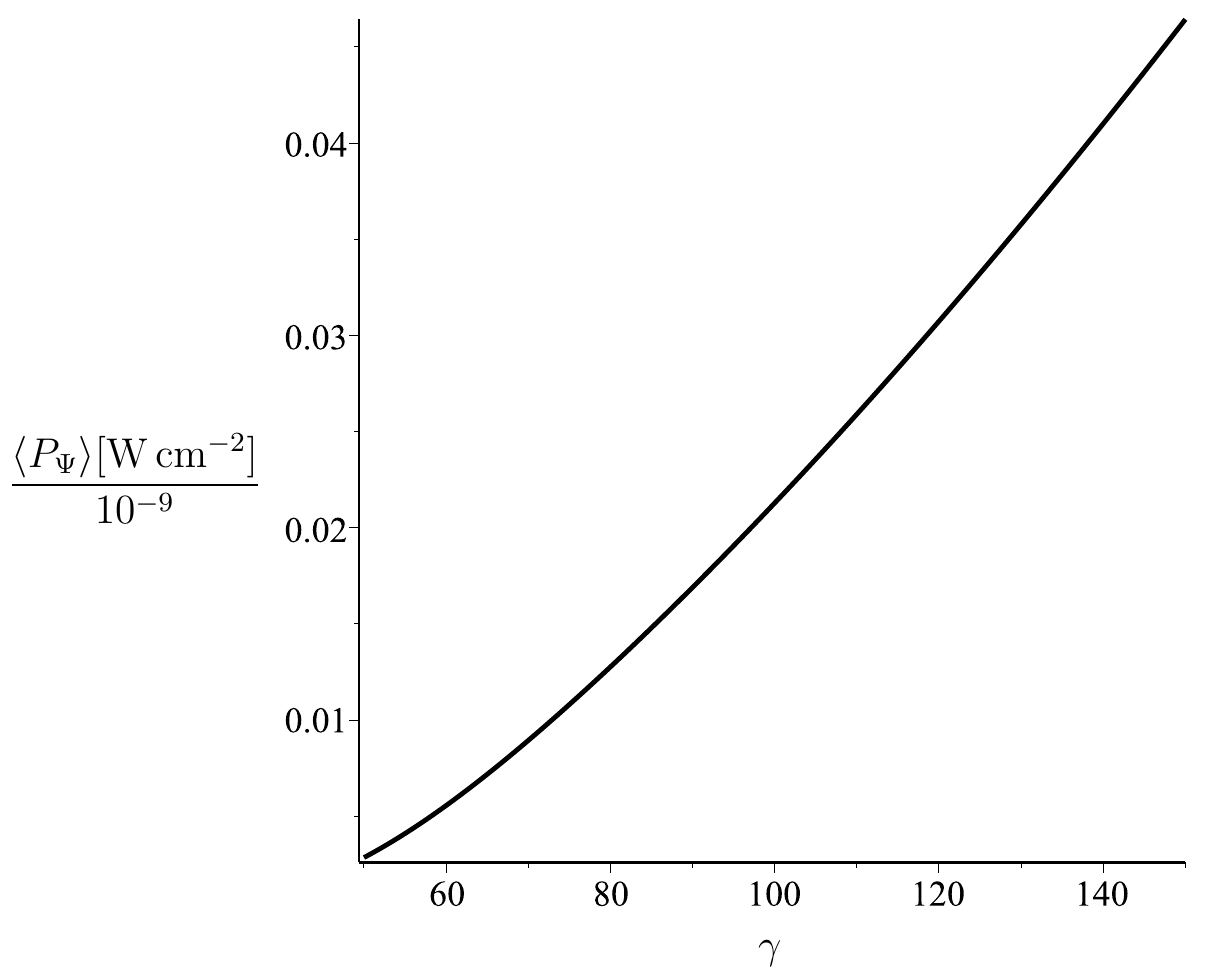}
\label{fig:P_vs_gamma_high_m}
\end{minipage}
}
\subfigure[]{
\begin{minipage}[t]{0.5\textwidth}
\centering\includegraphics[scale=0.6]{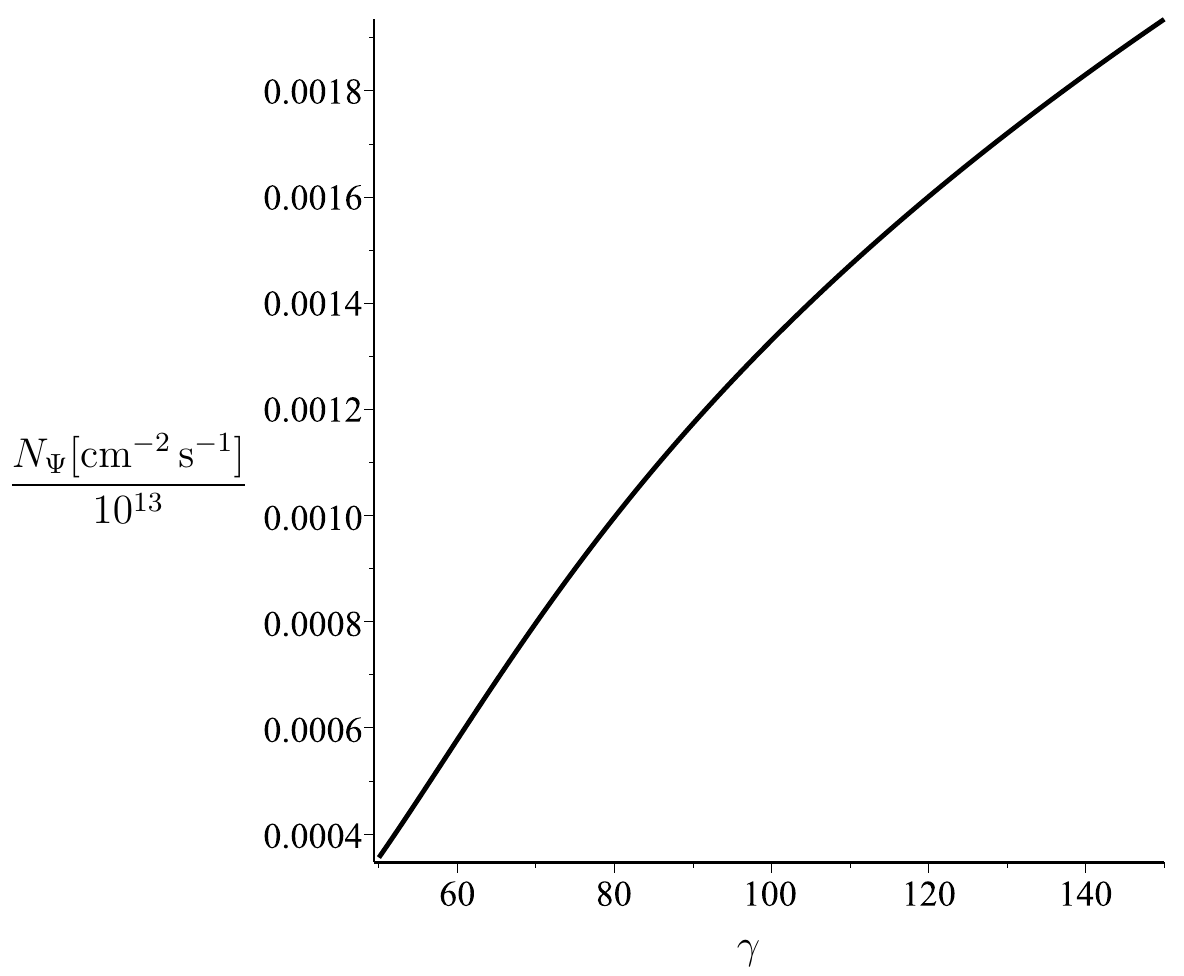}
\label{fig:N_vs_gamma_high_m}
\end{minipage}
}
{\color{black}\caption{\label{fig:fluxes_vs_gamma_high_m}The average ALP energy flux density (\ref{av_P_Psi_final}) and average ALP flux density (\ref{av_N_Psi_final}) versus the Lorentz factor $\gamma$ of the wake for the parameters $\omega_p = 2\pi\times 10^{13}\,{\rm rad}\,{\rm s}^{-1}$, $B = 35\,{\rm T}$, $g = 0.66 \times 10^{-10}\,{\rm GeV}^{-1}$, $m_\Psi = 10^{-4}\,{\rm eV}/c^2$.}}
\end{figure}

{\color{black} The above considerations are also applicable, as order-of-magnitude estimates, to the 3-dimensional bubble, or blow-out, regime, which features prominently in laser-driven plasma wakefield acceleration. Analytical considerations and PIC simulations demonstrate that the electric field in the bubble regime is $\sim 50\%$ of the electric field in a 1-dimensional wake~\cite{kostyukov:2004}. Inspection of ({\ref{ALP_eqn}}), (\ref{ALP_energy_flux_density}) shows that $\Psi$ scales as $|{\bf E}|$, and $|{\bf P}_\Psi|$ scales as $\Psi^2$, respectively. Hence, the average flux density of ALPs produced in the bubble regime is expected to be $\sim 25\%$ of (\ref{av_N_Psi_final}).}
{\color{black} Furthermore, the dependence of $\langle P_\Psi \rangle$, $N_\Psi$ on the peak frequency $\omega_0$ of the laser pulse can be introduced using the relationship $\gamma \approx \omega_0/(\sqrt{3}\,\omega_p)$~\cite{lu:2007}. 

Figs. \ref{fig:fluxes_vs_gamma_low_m}, \ref{fig:fluxes_vs_gamma_high_m} show that the behaviour of $\langle P_\Psi \rangle$, $N_\Psi$ versus $\gamma$ is highly sensitive to the value of $m_\Psi$. Moreover, a small change in $\gamma$ for fixed $m_\Psi$ can induce relatively large changes in the properties of the ALP field $\Psi$. The precise details of the coefficient of proportionality in $\gamma \propto \omega_0/\omega_p$ are needed to reveal the dependence of $\langle P_\Psi \rangle$, $N_\Psi$ on $\omega_p$ for a specific laser pulse.}

Whilst the solar ALP flux is essentially continuous, from a practical perspective a laser-driven plasma wakefield accelerator can, at best, provide a pulsed source of ALPs because the plasma must be replenished between laser shots. Our estimates correspond to a single pulse of ALPs whose duration is expected to be tens of femtoseconds. However, using the forthcoming ELI facilities~\cite{eli}, it should be possible to produce a train of such pulses with a repetition rate of a few ${\rm Hz}$. 

\begin{figure}
\centering\includegraphics{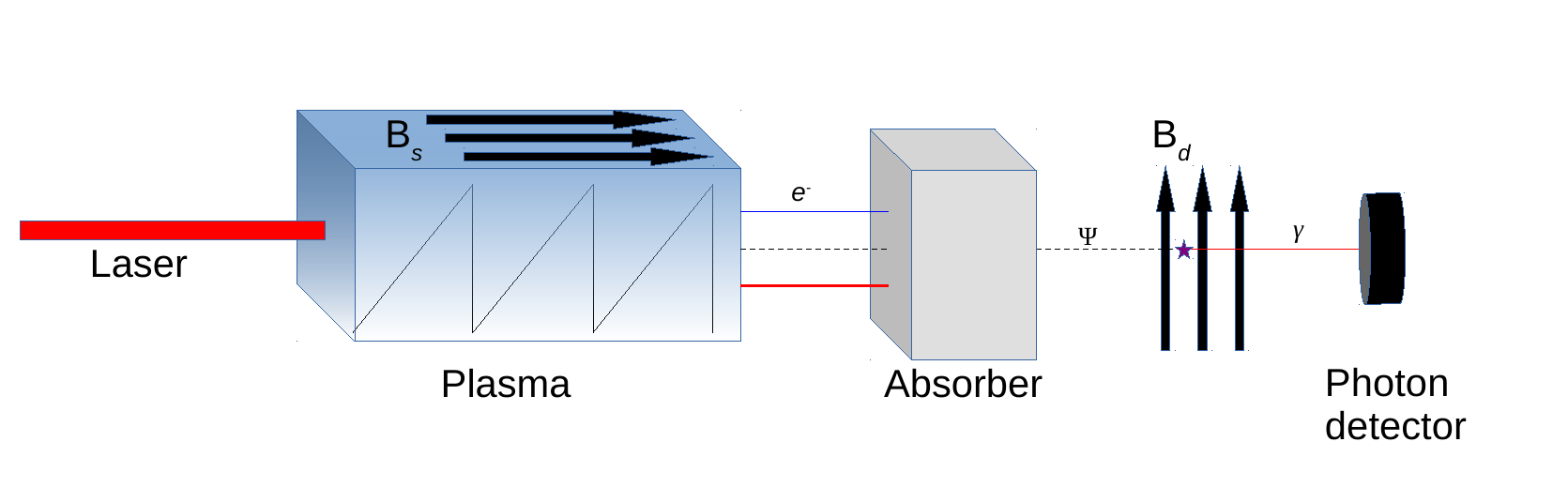}
\caption{\label{fig:schematic} {\color{black} A novel light-shining-through-wall experiment in which a laser-driven non-linear plasma wave (indicated by a sawtooth) interacts with a static longitudinal magnetic field to produce ALPs (dashed line), energetic electrons (blue line) and intense photons (thick red line). Photons, electrons and ALPs emanate from the laser-driven plasma. The photons and electrons are absorbed downstream, whilst the ALPs penetrate the absorber and are converted to ${\rm THz}$ photons (thin red line) using a static transverse dipole field.}}
\end{figure}

{\color{black} The plasma channel of a laser-wakefield accelerator could be sited within the bore of a powerful solenoid whose magnetic field is essentially uniform over hundreds of plasma wavelengths. For example, the change in the magnetic field away from the centre of a standard MagLab~\cite{NHMFL} $35\,{\rm T}$, $32\,{\rm mm}$ bore solenoid magnet is less than $\sim 1.0\%$ within $6\,{\rm mm}$ parallel to the bore axis and less than $\sim 0.4\%$ within $16\,{\rm mm}$ perpendicular to the bore axis. The effects of the ALPs emerging from the opposite end of the solenoid could be detected using apparatus similar to CAST, in which ALPs are converted to photons over the $9.26\,{\rm m}$ length of a $\sim 9\,{\rm T}$ dipole magnet~\cite{CAST:2017}. However, unlike solar ALPs which yield X-rays, (\ref{Psi_fourier}),(\ref{l_approx}) can be used to show that the photon frequencies would be integer multiples of $\sim 1\,{\rm THz}$. Clearly, the detector would need to be shielded from the intense radiation emitted by plasma electrons driven by the laser wakefield accelerator. Hence, one can regard such a set-up as an novel light-shining-through-wall experiment~{\color{black}\cite{redondo:2011}} in which one can manipulate the flux of ALPs by altering the current through the solenoid and the laser-plasma parameters. See Figure~\ref{fig:schematic}.}
\section{Conclusion}
The world-wide effort to identify the nature of dark matter is on-going, and axion-like particles (ALPs) are a promising candidate. Although most searches for ALPs rely on candidate astrophysical sources, there are practical advantages in locating both the candidate source and detector in a terrestrial laboratory. {\color{black} We have obtained an initial estimate of the flux density of ALPs generated by the wakefield trailing an intense laser pulse propagating through a plasma.} Our estimates suggest that a laser-driven plasma wakefield propagating along the strongest static magnetic field available in the laboratory generates a pulse of ALPs whose flux density is greater than the flux density of solar ALPs at Earth if the ALP mass is less than $\sim 10^{-4}\,{\rm eV}/c^2$. Furthermore, the average ALP energy flux density generated by the wakefield is close to its maximum theoretical value for an ALP mass less than $\sim 10^{-5}\,{\rm eV}/c^2$, which is in the mass range of state-of-the-art experiments that use microwave cavities to search for dark matter. {\color{black} The above results naturally suggest a new genre of light-shining-through-wall experiments.} Taken together, these {\color{black} considerations} lead us to conclude that detailed investigations should be undertaken of the benefits offered by laser-driven plasma wakefields in contemporary searches for ALPs.
\section*{Acknowledgements}
This work was supported by the UK Engineering and Physical Sciences Research Council grant EP/N028694/1. We thank Ian R Bailey and Christopher T Hill for useful discussions, and we thank the referees for their useful comments. All of the results can be fully reproduced using the methods described in the paper.


\begin{thebibliography}{99}
\bibitem{eli} 
http://www.eli-beams.eu/
\bibitem{dipiazza:2012} 
A. Di Piazza, C. M\"{u}ller, K.Z. Hatsagortsyan, C.H. Keitel, Rev. Mod. Phys. 84 (2012) 1177.
\bibitem{burton:2014} 
D.A. Burton, A. Noble, Contemp. Phys. 55 (2014) 110.
\bibitem{mendonca:2007} 
J.T. Mendon\c ca, EPL 79 (2007) 21001.
\bibitem{dobrich:2010} 
B. D\"obrich, H. Gies, JHEP 10 (2010) 022.
\bibitem{villalba:2013} 
S. Villalba-Ch\'avez, A. Di Piazza, JHEP 11 (2013) 136.
\bibitem{villalba:2014} 
S. Villalba-Ch\'avez, Nucl. Phys. B 881 (2014) 391.
\bibitem{peccei:1977} 
R.D. Peccei, H.R. Quinn,  Phys. Rev. Lett. 38 25 (1977) 1440.
\bibitem {weinberg:1978} 
S. Weinberg, Phys. Rev. Lett. 40 4 (1978) 223.
\bibitem{wilczek:1978} 
F. Wilczek,  Phys. Rev. Lett. 40 5 (1978) 279.
\bibitem{cicoli:2012} 
M. Cicoli, M.D. Goodsell, A. Ringwald, JHEP 10 (2012) 146.
\bibitem{preskill:1983} 
J. Preskill, M.B. Wise, F. Wilczek, Phys. Lett. B 120 1-3 (1983) 127.
\bibitem{abbott:1983} 
L.F. Abbott, P. Sikivie, Phys. Lett. B 120 1-3 (1983) 133.
\bibitem{dine:1983} 
M. Dine, W. Fischler, Phys. Lett. B 120 1-3 (1983) 137.
\bibitem{baker:2013} 
K. Baker, {\it et al.}, Ann. Phys. (Berlin) 525 6 (2013) A93.
\bibitem{rosenberg:2015} 
L.J. Rosenberg, PNAS 112 40 (2015) 12278.
\bibitem{seviour:2014} 
R. Seviour, I. Bailey, N. Woollett, P. Williams, J. Phys. G: Nucl. Part. Phys. 41 (2014) 035005.
\bibitem{caldwell:2017} 
A. Caldwell {\it et al.}, Phys. Rev. Lett. 118 (2017) 091801. 
\bibitem{tajima:1979} 
T. Tajima, J.M. Dawson, Phys. Rev. Lett. 43 4 (1979) 267.
\bibitem{mangles:2004} 
S.P.D. Mangles, {\it et al.}, Nature 431 (2004) 535.
\bibitem{geddes:2004} 
C.G.R. Geddes, {\it et al.} Nature 431 (2004) 538.
\bibitem{faure:2004} 
J. Faure, {\it et al.} Nature 431 (2004) 541.
\bibitem{schlenvoigt:2008} 
H.P. Schlenvoigt {\it et. al.}, Nat. Phys. (2008) 4 130.
\bibitem{blumenfeld:2007} 
I. Blumenfeld {\it et. al.}, Nature 445 (2007) 741.
\bibitem{assman:2014} 
R. Assman {\it et al.}, Plasma Phys. Control. Fusion 56 (2014) 084013.
\bibitem{esarey:2009} 
E. Esarey, C.B. Schroeder, W.P. Leemans, Rev. Mod. Phys. (2009) 1229.
\bibitem{pukhov:2002} 
A. Pukhov, J. Meyer-Ter-Vehn, Appl. Phys. B 74 (2002) 355. 
\bibitem{lu:2007} 
W. Lu {\it et al.}, Phys. Rev. ST Accel. Beams 10 (2007) 061301.
\bibitem{marklund:2006} 
M. Marklund, P.K. Shukla, Rev. Mod. Phys. 78 (2006) 591.
\bibitem{born:1934} 
M. Born, L. Infeld, Proc. Roy. Soc. A 144 (1934) 425.
\bibitem{fradkin:1985} 
E.S. Fradkin, A.A. Tseytlin, Phys. Lett. B 163 (1985) 123.
\bibitem{burton:2010} 
D.A. Burton, A. Noble, J. Phys. A: Math. Theor. 43 (2010) 075502.
\bibitem{thomas:2016} 
A.G.R. Thomas, Phys. Rev. E 94 (2016) 053204.
\bibitem{burton:2016} 
D.A. Burton, A. Noble, T.J. Walton, J. Phys. A: Math. Theor. 49 (2016) 385501.
\bibitem{burton:2011} 
D.A. Burton, R.M.G.M. Trines, T.J. Walton, H. Wen, J. Phys. A: Math. Theor. 44 (2011) 095501.
\bibitem{schroeder:2011} 
C.B. Schroeder, C. Benedetti, E. Esarey, W.P. Leemans, Phys. Rev. Lett. 106 (2011) 135002.
\bibitem{NHMFL} 
http://nationalmaglab.org/
\bibitem{CAST:2017} 
V. Anastassopoulos {\it et. al.}, Nat. Phys. 13 (2017) 584.
\bibitem{andriamonje:2007} 
S. Andriamonje {\it et al.}, JCAP 04 (2007) 010.
\bibitem{kostyukov:2004} 
I. Kostyukov, A. Pukhov, S. Kiselev, Phys. Plasmas 11 (2004) 5256.
\bibitem{redondo:2011} 
J. Redondo, A. Ringwald, Contemp. Phys. 52 (2011) 211.
\end{thebibliography}
\end{document}